\documentclass[conference,a4paper]{IEEEtran}
\PassOptionsToPackage{draft}{hyperref}

\usepackage{graphicx}
\usepackage{cite}
\usepackage[cmex10]{amsmath}
\usepackage{amssymb}
\usepackage{url}
\usepackage[caption=false,font=footnotesize]{subfig}
\usepackage{fixltx2e}
\usepackage{stfloats}
\usepackage[usenames,dvipsnames]{color}
\usepackage{balance}
\usepackage{algorithm}
\usepackage{algorithmicx}
\usepackage{algpseudocode}

\begin{document}
\bstctlcite{IEEEexample:BSTcontrol}
\title{Spatial Coordination Strategies in\\Future Ultra-Dense Wireless Networks}

\author{\IEEEauthorblockN{Antonis G. Gotsis, Stelios Stefanatos, and Angeliki Alexiou}\\
\IEEEauthorblockA{Department of Digital Systems, University of Piraeus, Greece\\
126 Grigoriou Lampraki Street, Office 501, GR18532, Piraeus, Greece\\
Email: \{agotsis,sstefanatos,alexiou\}@unipi.gr}}


\maketitle

\begin{abstract}
\boldmath
Ultra network densification is considered a major trend in the evolution of cellular networks, due to its ability to bring the network closer to the user side and reuse resources to the maximum extent. In this paper we explore spatial resources coordination as a key empowering technology for next generation (5G) ultra-dense networks. We propose an optimization framework for flexibly associating system users with a densely deployed network of access nodes, opting for the exploitation of densification and the control of overhead signaling. Combined with spatial precoding processing strategies, we design network resources management strategies reflecting various features, namely local vs global channel state information knowledge exploitation, centralized vs distributed implementation, and non-cooperative vs joint multi-node data processing. We apply these strategies to future UDN setups, and explore the impact of critical network parameters, that is, the densification levels of users and access nodes as well as the power budget constraints, to users performance. We demonstrate that spatial resources coordination is a key factor for capitalizing on the gains of ultra dense network deployments.
\end{abstract}


% \fi
\IEEEpeerreviewmaketitle

%5G Radio Access; Ultra-Dense Wireless Networks; Spatial Resources Coordination; Pairing; Precoding; Optimization
%5G Radio Access; Small-Cell Networks; Network Densification; Radio Resources Management; Coordination; Optimization; Integer Linear Programming; Pairing; Partitioning; Power Control

\section{Introduction}\label{sec:Intro}

\subsection{Motivation}\label{sec:Intro_Mot}
The rapid worldwide deployment of HSPA (3.5G) and LTE/LTE-A (4G) networks~\cite{4GAmerNetdeploymentApr14} marks the beginning of the true mobile broadband era. Several Mbps data rates are now offered through state-of-the-art cellular infrastructures. At the same time, the predicted traffic load requirements for the not-so-distant future, call for the evolution of current radio access networks towards a new generation of technology arriving by 2020. 5G is expected to deliver in an energy-efficient way, up to three orders of magnitude higher network capacity as well as scalable quality-of-service (QoS) levels on a per user (UE) basis~\cite{METISD62}. Despite the fact that 5G is in its infancy since even use-cases, scenarios, requirements and key-performance-indicators have not been decided yet, the telecom world, involving both industry and academia, are highly eager to conceptualize 5G and develop relevant technical solutions. Collaborative research projects and fora~\cite{METISD62}, industry~\cite{NSN5GWP13} and standardization groups efforts~\cite{3GPPR12feb14,HEW} highlight this trend.

Along these lines, extreme network densification empowered with universal resources reuse capabilities is considered one of the strongest 5G themes~\cite{BhJu14}. Although the concept of cell-splitting/small-cells is nothing new, ultra-dense networks (UDNs) aspire to go beyond the cellular offloading use-case through rethinking of the current radio network practices. To cope with the multi-fold increase in UE population and demanding rates per UE, the typical area used to be covered by a macro-cell, should be populated with tens or even hundreds of low-power infastructure/access nodes (ANs)~\cite{GeLe13} (such as small-cells, distributed antennas or access points). Different from cellular networks, UDNs are characterized by comparable and even higher infrastructure elements density (including nodes and antennas per node) with that of UEs\cite{METISD62}. Such dense network deployments are hardly regular, creating complex and unpredictable interference conditions. On top of that, UE distributions are highly inhomogeneous, while QoS should be guaranteed on a per-UE basis. Therefore, ultra network densification and universally coordinated resources management should go hand-in-hand, in order to achieve scalable capacities in future ever-densified radio access networks~\cite{Al14,LiNi14}.

From a system and standardization evolution point of view, the UDN theme is considered a key future 3GPP feature (for the upcoming release 12 and onwards)~\cite{3GPP_36872}. Spatial coordination techniques (referred to as ``CoMP" in the LTE community), such as coordinated scheduling/beamforming and joint processing have been recently examined and evaluated, under the future envisioned dense-small cell topology layer~\cite{3GPP_36874}. In addition, UDN comprises the central pillar of the recently approved IEEE 802.11ax project (previously known as High-Efficiency WiFi or ``HEW") which is expected to commence by June 2014 and aims at defining the next evolution wave for the WLAN technology~\cite{HEW}.

This paper aims at shedding some light on the design and rate performance of coordinated spatial resources management strategies for future 5G UDNs. In this context, we study techniques for associating each UE with one or more multi-antenna ANs (termed AN-UE \emph{pairing} hereafter), as well as obtaining efficient network precoding matrices (involving beamforming weight vectors and power-scaling factors), and explore their performance in terms of worse-UE supported rate. Practical system limitations imposed by signaling and backhaul requirements, e.g. for reporting the channel-state-information and the resource allocation decisions, as well as exchanging users' data among ANs, are further exacerbated in UDNs, questioning the efficient network scalability. To account for these issues we consider and contrast various operation and resources allocation optimization modes, including local (per-AN) optimization, fully-centralized or distributed coordination in the signaling plane, as well as cooperation in the data plane. By applying these alternatives to various random dense deployment scenarios, we explore rate scaling trends and obtain useful insights for future UDN design and parametrization.

\subsection{Related Work}\label{sec:Intro_RelWork}
Focusing on requirements and technology solutions for future high capacity 5G networks, the works in~\cite{GeLe13}~and~\cite{HwSo13} explored the densification requirements from a simulation (3GPP-based) and an analytical (stochastic-geometry based) perspective. However, no intelligent coordination mechanisms were considered, and the quantification of area throughput was targeted, which fails on capturing the discrepancies in experienced QoS among system UEs. In~\cite{LiWu13} the average spectrum efficiency for UDNs was analyzed leveraging stochastic geometry and the impact of parameters such as access nodes and UEs density and transmit power was explored; however, the spatial dimension exploitation through multi-antenna and multi-AN coordinated/coopeative processing was not taken into account. In~\cite{GoAl13pimrc} joint pairing and power coordination optimization strategies for dense network deployments were developed, but only for single-antenna small-cells, while in~\cite{GoAl13gc} optimal and suboptimal pairing and precoding mechanisms were proposed, but mainly focusing on backhaul requirements reduction for small-scale distributed antenna networks.

\subsection{Contribution and Paper Organization}\label{sec:Intro_Contr}
By recognizing that coordination in future ultra-dense networks is of paramount importance and opting for the provision of scalable QoS levels on a per-UE (and not only on a per-area) basis, we propose and evaluate various relevant spatial resources management strategies. In particular, a multi-UE/AN pairing optimization framework is defined in a unified way, as the key pillar for enabling various local, coordination and cooperation precoding in UDNs. A mathematical programming formulation for the AN-UE pairing problem is devised, reflecting system overhead limitations (for exchanging UEs' data among coopeative ANs) and providing different opportunities for spatial resources (beams) sharing among UEs served by a single multi-antenna AN. We leverage integer linear programming (ILP) to solve the above problem, and as a subsequent step well-known network precoding algorithms for controlling interference and maximizing the worse-UE rate performance are applied. A detailed comparative performance evaluation study of the algorithmic approaches is performed, examining different power budgets as well as AN and UE density scenarios, starting from the typical cellular up to the ultra-dense region with excessive number of infrastructure elements to UEs. We finally discuss the pros and cons of each strategy in terms of achievable performance and implementation/overhead requirements, and demonstrate that in the ultra-dense region the determination of the best strategy to follow depends on both AN/UE density and power conditions.

The remaining of this paper is structured as follows. Sec.\ref{sec:SystemModel} introduces the system model along with the basic assumptions and parameters. Sec.\ref{sec:CoordStrategies} defines the pairing optimization framework, provides the problem formulation and solution, and addresses how this framework supports alternative resources management algorithms operating on a local, coordination or cooperation mode. Performance evaluation results as extracted through Monte-Carlo simulations are given in Sec.\ref{sec:Results}, along with discussion on key findings and insights. The paper is concluded in Sec.\ref{sec:Conclusion}, where also open issues and potential future work items are proposed.

\section{System Model and Problem Description}\label{sec:SystemModel}
\subsection{System Model and Assumptions}
We assume a typical macro-cell area covered by a random topology network of $M$ access nodes or ANs (where $\mathcal{M}$ is the ANs set), each equipped with $L$ antennas, serving $K$ single-antenna users or UEs (where $\mathcal{K}$ is the UEs set). The bandwidth is fully reused at each AN and no orthogonal partitioning scheme (e.g. TDMA/FDMA) is applied; thus we examine a single unit resource element, such as an LTE time-frequency resource block, considered to be spatially reused across all UEs. Along these lines, all UEs are served simultaneously, hence the network infrastructure consists of at least ${M^{{\text{min}}}} = \left\lceil {{K \mathord{\left/
 {\vphantom {K L}} \right.
 \kern-\nulldelimiterspace} L}} \right\rceil$ ANs, such that for each UE at least one infrastructure element or spatial degree of freedom (dof) is available. For a UDN the density of access nodes (${\lambda _{{\text{AN}}}}$) is comparable and even surpasses that of UEs (${\lambda _{{\text{UE}}}}$), where each density level is calculated as the number of nodes lying in the given area.

Following the notation of~\cite{BjJo13}, we define the $L$-length channel state information (CSI) vector regarding an arbitrary communications pair formed by the $k^{\text{th}}$ UE and the $m^{\text{th}}$ AN as ${{\mathbf{h}}_{mk}} \in {{\mathbb{C}}^{L \times 1}}$. Then, the complete CSI vector for the particular user can be acquired by stacking all the per-AN vectors as ${{\mathbf{h}}_k} = {\left[ {{\mathbf{h}}_{1k}^T \ldots {\mathbf{h}}_{Mk}^T} \right]^T} \in {{\mathbb{C}}^{ML \times 1}}$. Each CSI element captures both large- (path-loss) and small-scale (fading) propagation impact, as well as the effect of thermal noise at the receiver, by proper normalization. In particular, the arbitrary CSI element for the $l^\text{th}$ antenna element is given by $h_{km}^l = \sqrt {{{{g_{km}}} \mathord{\left/
 {\vphantom {{{g_{km}}} {{\sigma ^2}}}} \right.
 \kern-\nulldelimiterspace} {{\sigma ^2}}}}  \cdot \delta _{km}^l$, where ${{g_{km}}}$ is the path-gain factor (depending on ANs topology and UEs distribution), $\sigma ^2$ is the noise level power, and $\delta _{km}^l$ is the small-scale fading amplitude. Accordingly, we can define the precoding vector for user $k$ as ${{\mathbf{w}}_k} = {\left[ {{\mathbf{w}}_{k1}^T \ldots {\mathbf{w}}_{kM}^T} \right]^T} \in {{\mathbb{C}}^{ML \times 1}},{{\mathbf{w}}_{km}} \in {{\mathbb{C}}^{L \times 1}}$, capturing both beam-weights and power information. Then the signal to interference and noise ratio (SINR) on a per UE basis is given by:
${\gamma _k} = {{{{\left| {{\mathbf{h}}_k^H{{\mathbf{w}}_k}} \right|}^2}} \mathord{\left/
 {\vphantom {{{{\left| {{\mathbf{h}}_k^H{{\mathbf{w}}_k}} \right|}^2}} {\left( {1 + \sum\limits_{i \ne k} {{{\left| {{\mathbf{h}}_k^H{{\mathbf{w}}_i}} \right|}^2}} } \right)}}} \right.
 \kern-\nulldelimiterspace} {\left( {1 + \sum\limits_{i \ne k} {{{\left| {{\mathbf{h}}_k^H{{\mathbf{w}}_i}} \right|}^2}} } \right)}}.$ Utilizing a simple Shannon-based PHY abstraction model, we obtain the rate per UE as ${r_k} = {\log _2}\left( {1 + {\gamma _k}} \right)$. Note that the UE precoding vectors implicitly contain the pairing information; for example, if user $k$ is not served by the node ${m'}$, then ${{\mathbf{w}}_{m'k}} = {{\mathbf{0}}_{L \times 1}}$ holds.

 As far as the power budget constraints definition is concerned, we follow a two-level approach: (a) we first assume a total (across the network) maximum power budget ${{p_\text{tot}}}$ which is preserved independently from the AN and/or UE density scenario, and (b) we impose a per-AN maximum allocated power constraint, by evenly splitting $p_{\text{tot}}$ among only the active ANs. Note that ANs not assigned to any UE (inactive ANs) are turned-off. If we denote by $\mathcal{A}$ the set of active ANs, then the allocated power per AN $p_m$ is bounded by ${{{p_m} \leqslant {p_{tot}}} \mathord{\left/
 {\vphantom {{{p_m} \leqslant {p_\text{tot}}} {\left| \mathcal{A} \right|}}} \right.
 \kern-\nulldelimiterspace} {\left| \mathcal{A} \right|}},\forall m \in \mathcal{A}$. The constant power assumption is justified by the fact that future 5G networks should attain 10-100x energy efficiency improvements compared to today's networks~\cite{METISD62}. Given that data rates should be drastically enhanced by two-three orders of magnitude, we keep the overall transmission power, $p_\text{tot}$, constant irrespective of heavier deployment densification and/or larger UE population.

\subsection{Problem Description}
We aim at proposing, evaluating, and comparing for various UDN deployment scenarios, spatial resources management strategies which provide highly balanced rate levels among UEs, in other words optimize the worse-UE performance. This could be ideally achieved by selecting the AN-UE pairs (where, in general, one UE could be served cooperatively by multiple-ANs) and the precoding vectors, using a joint optimization procedure. However, as extensively discussed in~\cite{GoAl13gc}, this corresponds to a mixed-integer non-linear (MINLP) program, for which its complexity scales exponentially with the number of variables. For UDN setups comprising tens or even hundreds of infrastructure elements (nodes and antennas/node), acquiring good solutions in reasonable time is practically impossible. To cope with this limitation in this paper we follow a decoupled approach.

First, we address the \emph{pairing optimization} sub-problem ignoring precoding. We seek for determining the optimal AN-UE pairs exploiting only the large-scale channel state information, expressed through the path-gain factors $\left\{ {{g_{km}}} \right\}$. Besides reducing complexity, this is a reasonable approach, since consideration of small-scale channel variations may lead to highly frequent changes in connectivity patterns (handovers) and to excessive signaling. Instead, the approach followed here allows for reconfiguring the pairing decision at a longer time-scale. However this remains a difficult combinatorial optimization problem. Secondly, given a pairing, \emph{linear precoding} is performed aiming at controlling interference and providing the required balanced QoS levels. Note that precoding optimization admits a global solution~\cite{ToPe09}, based on solving a series of (convex) second-order cone programming (SOCP) problems. One merit of introducing the particular pairing framework is its generality, since it can capture various operation alternatives, such as local (per-AN) optimization, coordinated precoding (exchange of signaling information) and joint processing (exchange of signaling and data)\footnote{For a brief analysis of the spatial precoding optimization methodology refer to the Appendix.}. Conclusively, in this paper we focus on the pairing optimization problem, while being aware of the underlying precoding alternatives.

\section{Spatial Coordination Strategies}\label{sec:CoordStrategies}

\subsection{The Pairing Optimization Framework}\label{sec:CoordStrategies_Pairing}

In a future UDN, a variety of potential serving ANs lie at the vicinity of each UE. Going beyond the cellular paradigm where the association rule for each UE is typically based only on the proximity of the serving AN, multiple additional criteria, such as propagation, traffic loading per AN and energy considerations (e.g. load shifting for turning off an AN) become relevant in UDNs~\cite{HoRa14,3GPP_36872}. In this context, we define a \emph{pairing} problem as the procedure for selecting the serving AN subset (among all the available ANs) for all system UEs. We provide a generic problem formulation admitting several types of constraints related to the utilization of available spatial degrees of freedom (dof), signaling and backhaul requirements. Since we assume that pairing is based on large-scale channel conditions we drop the antenna element index $l$.

We define a set of $KM$ binary UE-to-AN association variables $\left\{ {{\rho _{km}}} \right\}$, with $\rho_{km}=1$ if and only if the $k^{\text{th}}$ UE is associated with the $m^{\text{th}}$ AN. We also define a set of $M$ (auxiliary) binary variables $\left\{ {{\alpha _m}} \right\}$, with $a_m=1$ if and only if the $m^{\text{th}}$ AN serves at least one UE. In such case we say that this AN belongs to the subset of ``Active ANs" denoted by $\mathcal{A}$, where $\mathcal{A} \triangleq \bigcup {\left\{ {m \in \mathcal{M}:{a_m} = 1} \right\}}$. For each UE $k$ we define the single serving AN $\mathcal{S}_k$/ For non-cooperative operation modes (local or coordinated) the serving AN is the sole AN carrying each UEs data. On the contrary, for cooperative operation modes (joint processing) where each UEs data is offered by multiple ANs, the serving AN is defined as the responsible AN (or ``master" AN) for controlling the cooperative transmission parameters and forwarding data to the assisting ANs. In order to provide a fairly generic problem formulation we introduce the following parameters:
\paragraph{$\left\{ {{c_{km}}} \right\}$}The costs of associating each UE with each AN; they could be based on the AN-UE distances, path-gains or any function of them.
\paragraph{$b_{\text{max}}$} The maximum number of active ANs  (equivalently, the cardinality of set $\mathcal{A}$) serving all the system UEs; this could be dictated by overhead constraints. The required overhead could be limited through reduced-cardinality active ANs subsets, since with a full set $\mathcal{A}$: i) each UE should report back to each AN the CSI, ii) for cooperative transmissions, the payload data per UE should be forwarded from the Serving AN to all the assisting ANs. Clearly, $\left\lceil {{K \mathord{\left/
 {\vphantom {K L}} \right.
 \kern-\nulldelimiterspace} L}} \right\rceil  \leqslant {b_{\max }} \leqslant K$, where the lower bound guarantees that there are enough spatial dof to serve all UEs, and the upper bound provides a fair comparison for coordinated and cooperative modes.
\paragraph{$u_{\text{max}}$} The maximum number of UEs sharing the same Serving AN, selected from the range $1 \leqslant {u_{\max }} \leqslant L$. The upper bound guarantees that intra-AN interference can be totally eliminated (through e.g., zero-forcing precoding), whereas the lower bound corresponds to the case of a UE not sharing its Serving AN with another UE.

We may now provide the mathematical programming representation of the pairing problem as follows:
\begin{subequations}\label{eq:PairingILP}
\begin{gather}
  \mathop {\min }\limits_{\left\{ {{\rho _{km}}} \right\},\left\{ {{\alpha _m}} \right\}} \sum\limits_k {\sum\limits_m {{c_{km}} \cdot {\rho _{km}}} }  \hfill \label{eq:PairingILP_a}\\
  {\text{subject to}} \hfill \nonumber \\
  \sum\limits_m {{\rho _{km}}}  = 1,\forall k,{\rho _{km}} \leqslant {\alpha _m},\forall k,\forall m \hfill \label{eq:PairingILP_b}\\
  \sum\limits_m {{\alpha _m}}  \leqslant {b_{{\text{max}}}},\sum\limits_k {{\rho _{km}}}  \leqslant {u_{{\text{max}}}},\forall m \hfill \label{eq:PairingILP_c}
\end{gather}
\end{subequations}
The objective function in \eqref{eq:PairingILP_a} minimizes the cumulative cost for the selected AN-UE association decision. Note that alternatively we could minimize the maximum cost among all associations by considering:
\[\mathop {\text{min max} }\limits_{\left\{ {{\rho _{km}}} \right\},\left\{ {{\alpha _m}} \right\}} \left\{ {{c_{km}} \cdot {\rho _{km}}} \right\}{\text{ subject to \eqref{eq:PairingILP_b},\eqref{eq:PairingILP_c} }}.\]
However in this paper we stick with the initial objective function definition. In~\eqref{eq:PairingILP_b}, the first constraints guarantee that each UE is assigned to a single Serving AN, whereas the second constraints denote that a UE-to-AN association involving a non-active AN is not feasible. Finally, the two constraints subsets in~\eqref{eq:PairingILP_c} express the limits on the size of active ANs subset and number of UEs sharing a common Serving AN correspondingly. The optimization problem in~\eqref{eq:PairingILP_a}--\eqref{eq:PairingILP_c} belongs to the class of (binary) integer linear programs (ILP). For such problems complexity guarantees for acquiring the optimal solution are not provided. Nevertheless, there exists a series of powerful algorithms and solvers~\cite{ChBa10} which are able to provide the optimal solutions for even large-scale problem sizes (involving thousands of binary variables and hundreds of constraints) very fast\footnote{Note that optimal pairing could be straightforwardly found through exhaustive search, however this is computationally prohibitive due to involved problem sizes. In general for $M$ ANs and $K$ UEs complete enumeration requires $M^K$ combination checks, where $M$ and $K$ correspond to tens of nodes. On the contrary, for a typical example of 16 UEs and 32 ANs, the GUROBI ILP solver needs no more than 150 msec to return the optimal solution in a today's desktop computing platform.}.

\subsection{Spatial Resources Management Strategies}\label{sec:CoordStrategies_ALL}
Leveraging the pairing optimization framework introduced just above and the available precoding solutions, we define four alternative network resources management strategies with different features regarding: i) the amount of utilized small-scale CSI knowledge, namely local (CSI of each UE towards its Serving AN) or global (CSI of each UE towards all active ANs); ii) the network densification exploitation factor as dictated by the parameter $b_{\text{max}}$; note that this is directly associated with the backhaul requirements in our work; iii) the inter-AN interference management approach, namely, no inter-AN interference control (\emph{local} per-AN optimization), \emph{coordinated precoding} (centralized calculation of network precoding matrix but still each UE's data provided by a single AN), and \emph{cooperative precoding} where UEs' data are cooperatively combined by various ANs.

With respect to the definition of cost elements, we consider the inverse of large-scale path-gain, namely ${c_{km}} = g_{km}^{ - 1},\forall k,\forall m$, promoting proximal AN-UE pairs. In a UDN setup, such an approach, when combined with subsequent power coordination could lead to a drastic reduce of inter-pair interference and increased rate performance~\cite{Al14}. Regarding the parameter controlling the maximum number of co-AN UEs, $u_{\text{max}}$, we fix it to its upper bound $L$ (such that no intra-AN interference is present) and leave for future work a more evolved study\footnote{For coordinated precoding, allowing more UEs to be served by the same AN increases the possibility for a UE to be served by its closest AN but at the same time leaves less spatial dof to handle inter-AN interference.}.  Based on the above definitions we briefly describe the four considered strategies. Regarding pairing parametrization, for strategies (1)--(3) the number of active ANs is not limited, hence $b_\text{max} = K$, whereas for strategy (4) the lower bound of $\left\lceil {{K \mathord{\left/
 {\vphantom {K L}} \right.
 \kern-\nulldelimiterspace} L}} \right\rceil$ is used (as it will be further elaborated in what follows).

\subsubsection{\textbf{Local Precoding  (``Local")}} The first strategy performs uncoordinated precoding optimization (for example zero-forcing or SOCP-based) on a per-AN basis, utilizing only local-based CSI. Thus, inter-AN interference management is not supported. It serves as a baseline scheme, since it incurs no additional signaling load for reporting the global CSI and the precoding vectors decisions. This scheme reflects the minimum gains from increased deployment densification, as the network gets closer to the UE.
\subsubsection{\textbf{Coordinated Precoding (``CoordPr")}} Assuming that each UE can still be served by a single AN as above, the current scheme is based on the optimal precoding vectors determination in a centralized fashion. This approach requires global CSI knowledge and the calculation of the per-UE (equivalently per-AN) weights by a centralized entity. It adds to the densification gains provided by the "Local" scheme, by intelligently managing inter-AN interference. It provides the performance upper bound for any interference avoidance scheme which allows coordination and exchange of information only in the signaling plane.
\subsubsection{\textbf{Local Beamforming \& Power Coordination (``Local-PowCoord")}} The third strategy makes a compromise between the high achieved performance levels of ``CoordPr" and minimum overhead signaling requirements/low implementation complexity of ``Local". In particular, based on the same pairing solution as above, it first performs a per-AN beamforming optimization, and at a subsequent step performs a power scaling coordination on a per-UE basis utilizing the so-called effective CSI knowledge (which is simply given by ${\left| {{\mathbf{h}}_k^H{{\mathbf{w}}_k}} \right|^2}$, for each UE $k$). Details on the formulation and optimal solution of power coordination over a set of determined communication pairs can be found in \cite{TaCh13} and \cite{GoAl13pimrc}. Interestingly, the authors in \cite{TaCh13} showed that the exact optimal solution for power coordination can be obtained in a distributed fashion with modest message exchange; such features are really attractive from an implementation point of view. The performance of the current strategy is expected to lie in between the first two approaches, due to the decoupling of beamforming and power allocation sub-procedures.
\subsubsection{\textbf{Constrained Joint Processing (``JPcon")}} The last strategy introduces the ability of interference exploitation instead of interference avoidance as in approaches (2),(3). In other words, it allows the joint processing and transmission of all UEs' data by multiple distributed ANs, following the known ``Network MIMO" or, using the 3GPP terminology, JP-CoMP paradigm. Although cooperation provides a significant capacity boost, it is associated with highly challenging synchronization and backhaul requirements\cite{HuTu12}. To avoid excessive backhaul utilization, we apply a hard limit on the number of ANs that are able to cooperate for serving the system UEs. Therefore, we obtain the optimal solution for the problem in \eqref{eq:PairingILP} with ${b_{\max }}$ taken as $\left\lceil {{K \mathord{\left/
 {\vphantom {K L}} \right.
 \kern-\nulldelimiterspace} L}} \right\rceil$. Then, each UE should measure and report its small-scale CSI considering only the selected minimum cardinality active AN subset $\mathcal{A}$, and UEs' data should be also shared among only these ANs. Clearly, such an approach is expected to limit the densification exploitation gains by concentrating UEs to (potentially suboptimal) AN subsets in order to limit the size of active subset $\mathcal{A}$ (refer also to the extended version of this paper in~\cite{GoSe14} for a graphical illustration of this concept). At the same time, the performance benefits of JPcon are expected to stem from the fact that each UE is collaboratively served (through proper precoding weight vector selection) by multiple ANs.

 Summing up, the above strategies are designed to capture the trade-offs regarding achievable performance and the signaling/implementation complexity requirements, in an attempt to characterize the gains coming both from densification and advanced spatial resources management. In Table~\ref{ta:StratOverview} we provide a brief overview of the approaches, whereas in Figure~\ref{fig:RateVsDens} typical pairing optimization problem solutions for two settings of the $b_\text{max}$ parameter are graphically illustrated.

\begin{table*}[]
\footnotesize
\renewcommand{\arraystretch}{1.3}
\caption{Spatial Resources Management Strategies Overview}
\label{ta:Algs}
\centering
\begin{tabular}{p{1.8cm} p{1.2cm} p{14cm}}
\hline
\bfseries Strategy & \bfseries Pairing ($b_\text{max}$) & \bfseries Precoding (Beamforming and Power Control)\\
\hline\hline
{\color{red}\textbf{Local}} & $K$ & Per-AN Precoding Optimization: Zero-Forcing Multi-User MIMO due to its simplicity\\
{\color{NavyBlue}\textbf{CoordPr}} & $K$ & Global Coordinated Precoding Optimization through SOCP, assuming each UE's data available at a single AN\\
{\color{ForestGreen}\textbf{Local-PowCoord}} & $K$ & (Step 1) Per-AN Beamforming as in Local; (Step 2) Power Coordination utilizing the effective channels from Step 1\\
{\color{Black}\textbf{JPcon}} & $\left\lceil {{K \mathord{\left/ {\vphantom {K L}} \right.
 \kern-\nulldelimiterspace} L}} \right\rceil$ & Global Precoding Optimization assuming each UEs data available at all the ANs belonging to the Active ANs subset $\mathcal{A}$\\
\hline
\end{tabular}
\label{ta:StratOverview}
\end{table*}
\normalsize

\begin{figure*}[]
\centerline{
\subfloat[{Higher Exploitation of AN infrastructure: Strategies (1)--(3) }]
{\includegraphics[width=0.50\textwidth]{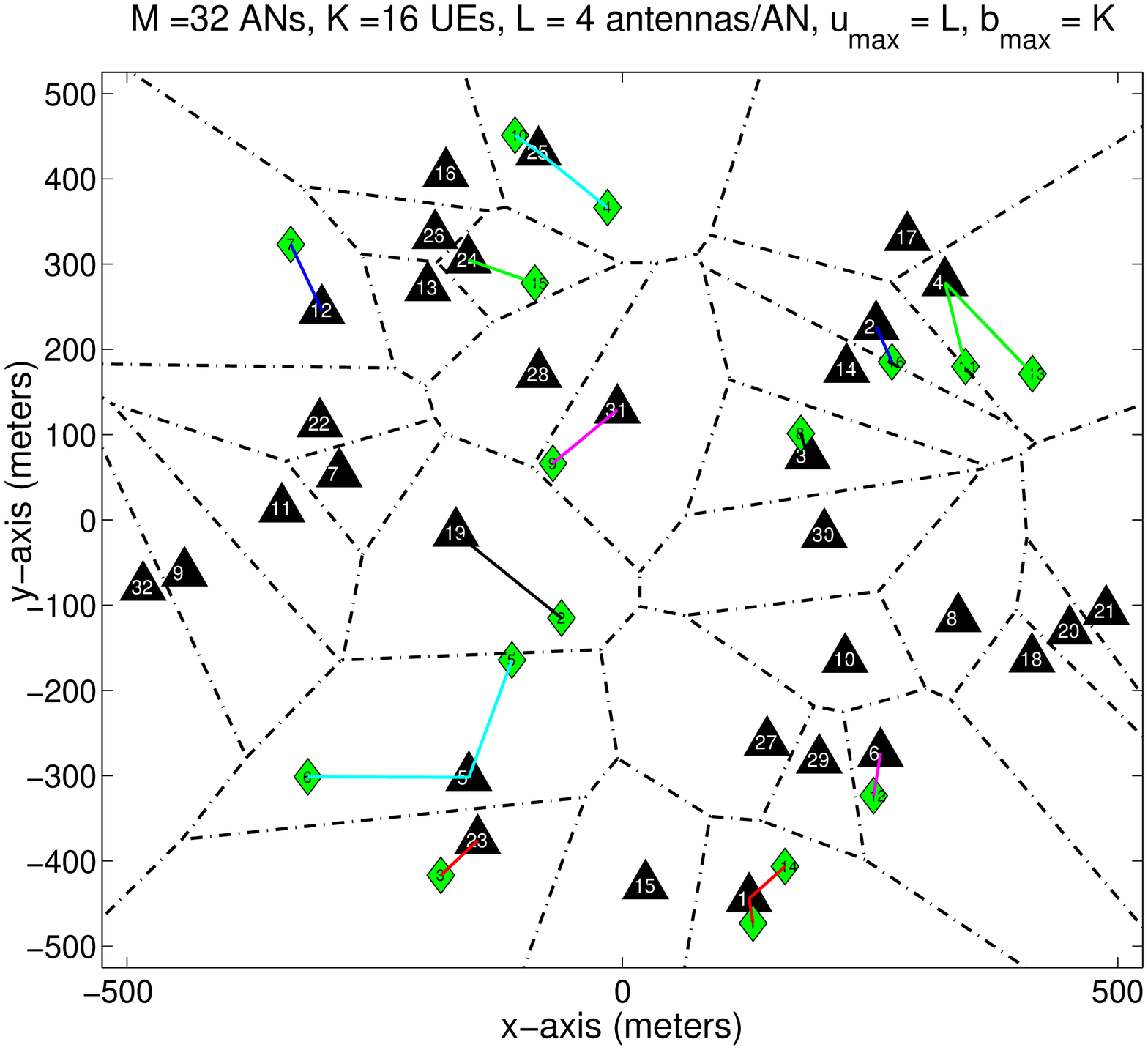}
\label{fig:RandomSnapshot_bmax_16}}
\hfill
\subfloat[{Limited Exploitation of AN infrastructure: Strategy (4)}]
{\includegraphics[width=0.50\textwidth]{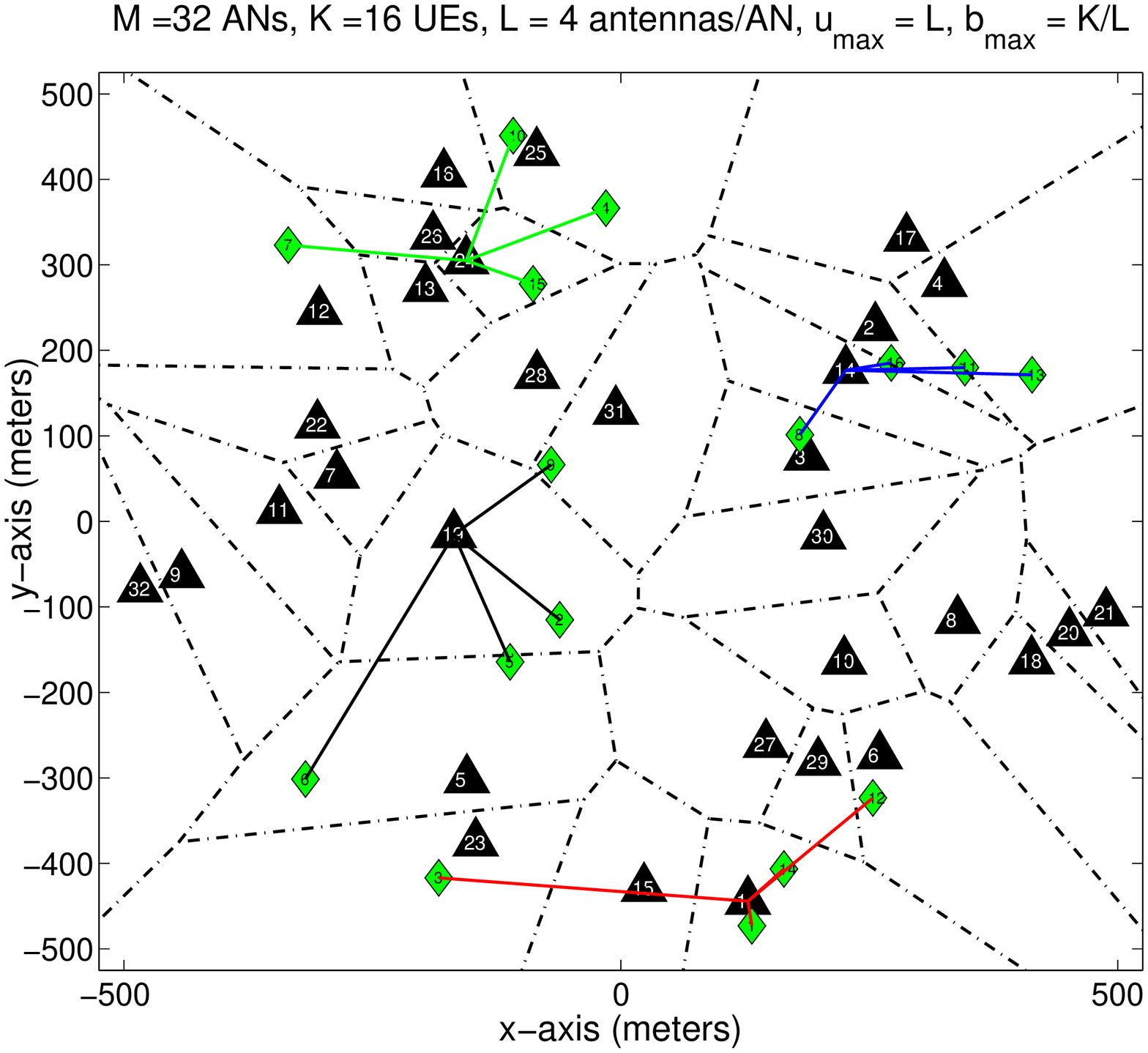}
\label{fig:RandomSnapshot_bmax_4}}
}
\caption{{Two pairing alternatives for a random network snapshot (The $\blacktriangle$ symbol stands for an AN, the $\Diamond$ symbol for a UE, and lines show the AN-UE pairs). On the left-hand side, $b_\text{max}$ is set to its upper bound value (16), hence the Active ANs subset ($\mathcal{A}$) has no cardinality limitation. Note that full densification exploitation allows for each UE to be paired with its closest AN (or voronoi cell~\cite{An13}). On the right-hand side, $b_\text{max}$ is set to its lower bound value (4), leading to reduced overhead requirements, but as shown, many UEs are forced to connect with a farthest AN, breaking the voronoi cells concept.}}
\label{fig:RateVsDens}
\end{figure*}

\section{Results \& Discussion}\label{sec:Results}
A set of indicative system scenarios is introduced, targeting the exploration of the impact of various critical factors, namely densification levels, degree of coordination and cooperation, as well as power availability, to the achieved performance as provided by each spatial resource management strategy. Each system scenario reflects a network deployment over an 1~$\text{km}^2$ square area, a given UE density $\lambda_\text{UE}$ equal to the number of randomly dropped single-antenna UE nodes $K$ in the 1~$\text{km}^2$ are), and a given AN density $\lambda_\text{AN}$ composed by $M$ ANs, equipped with $L=4$-antennas each, randomly dropped over the same area. A total power budget summed over and evenly split among all active ANs is also considered. Regarding propagation impact, a log-distance large-scale model with path-loss exponent equal to $4$ and a carrier frequency at 3.5 GHz is assumed, whereas for small-scale effects independent rayleigh fading over the spatial domain is considered. A thermal noise density level of -174 dBm/Hz is added to the received interference levels for each UE. Moreover, a single transmission resource unit spanning 180~kHz (similar to an LTE Resource Block) is reused among all ANs.

We evaluate through Monte-Carlo simulation, the supported worse-UE performance across the whole network (zero-outage) against the following two metrics: 1) the \emph{densification ratio} level, defined as ${{{\lambda _{{\text{AN}}}}} \mathord{\left/
 {\vphantom {{{\lambda _{{\text{AN}}}}} {{\lambda _{{\text{UE}}}}}}} \right.
 \kern-\nulldelimiterspace} {{\lambda _{{\text{UE}}}}}}$, and 2) the \emph{reference signal to noise ratio} $\text{SNR}_\text{ref}$, which is directly related to the total available power budget. $\text{SNR}_\text{ref}$ expresses the experienced worst-case received SNR for a user located at the area edge when it is served by a macro-BS located at the center. In this sense, we control network total power to achieve the desired $\text{SNR}_\text{ref}$ levels. For each scenario we simulate 250 statistically independent network snapshots and obtain the averaged worse-UE rate performance. For optimization modeling we use the CVX framework~\cite{CVX}, whereas for solving the ILP pairing problems we use GUROBI~\cite{GUROBIApr14}, and for the SOCP precoding problems, MOSEK~\cite{MOSEKApr14}.

\begin{figure*}[!t]
\centerline{
\subfloat[{Worse UE Rate Performance}]
{\includegraphics[width=0.50\textwidth]{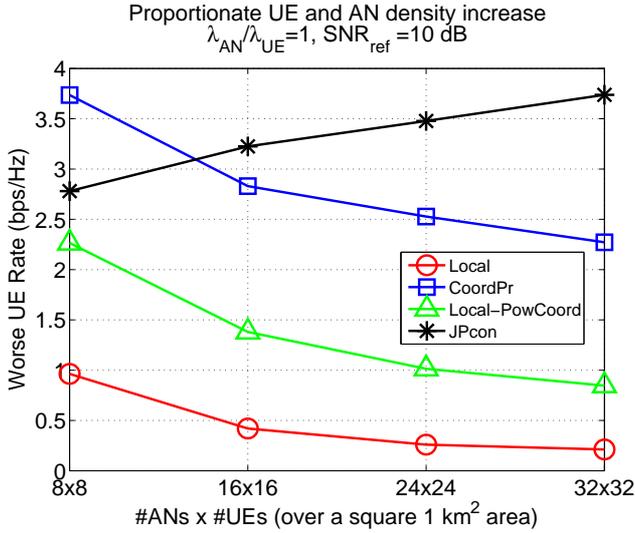}
\label{fig:SquareDens_WorseUERate}}
\hfill
\subfloat[{Network (Sum) Rate Performance for Uniform Rate Assignment}]
{\includegraphics[width=0.50\textwidth]{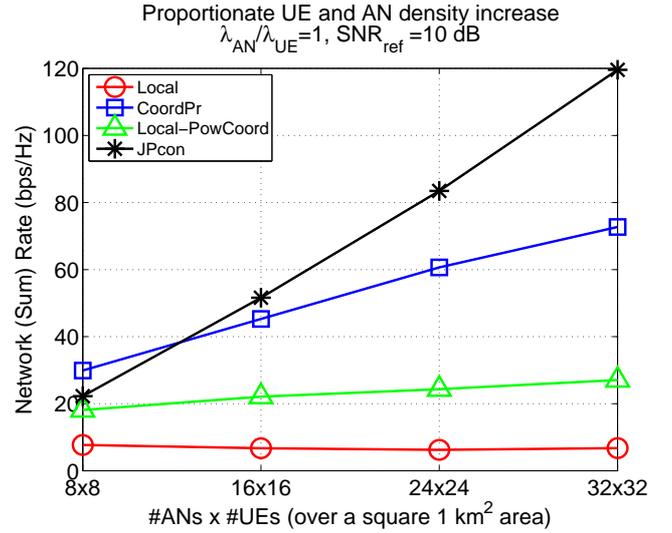}
\label{fig:SquareDens_SumRate}}
}
\caption{{Impact of Proportionate UE population increase and AN Densification on Worse-UE and Network Rate Performance}}
\label{fig:SquareDens}
\end{figure*}

\textbf{1) Impact of Proportionate UE population increase and AN Densification}: The first test scenario demonstrates the impact of ever-increasing UE population density. For a reference SNR level of 10 dB, we keep increasing the number of UEs requesting service in the same area and accordingly we densify the AN deployment, so as to preserve the densification ratio as ${{{\lambda _{{\text{AN}}}}} \mathord{\left/
 {\vphantom {{{\lambda _{{\text{AN}}}}} {{\lambda _{{\text{UE}}}}}}} \right.
 \kern-\nulldelimiterspace} {{\lambda _{{\text{UE}}}}}} = 1$. Figure~\ref{fig:SquareDens_WorseUERate} depicts the achieved worse-UE rate as access and user nodes increase. Compared to the baseline ``Local" strategy, both coordination strategies, namely ``Local−PowCoord" and ``CoordPr" significantly enhance the worse-UE performance by an average factor of $2-4\times$ and $4-10\times$ correspondingly. This is reasonable, since ever-densified deployments lead to more complex interference environments which limit the cell-edge performance levels. To address this issue, intelligently coordinating the selection of appropriate beamforming weights and power scaling, comes into play. However, as it is observed, no coordination strategy which aims at avoiding interference succeeds in sustaining a constant worse-UE rate level with a proportionate (1:1) AN/UE node density ratio.

 On the contrary, the cooperative strategy ``JPcon", not only preserves the worse-UE rate as the density of UEs becomes higher, but also enhances it; this is justified by the fact that it i) is able to exploit interference, rather than avoiding it, through joint network MIMO processing, and ii) exploits multi-UE and multi-AN diversity which becomes more prominent in larger-size networks. For low ANxUE population scenarios (8x8), ``JPcon" is outperformed by ``CoordPr" due to power limitation reasons. This will be also elaborated in the following discussion points. Finally, in Figure~\ref{fig:SquareDens_SumRate} we illustrate the corresponding sum-rate results, where the different rate-scaling slopes are clearly shown.

\begin{figure*}[!t]
\centerline{
\subfloat[{Impact of Densification Ratio and Power Budget for constant UE density}]
{\includegraphics[width=0.50\textwidth]{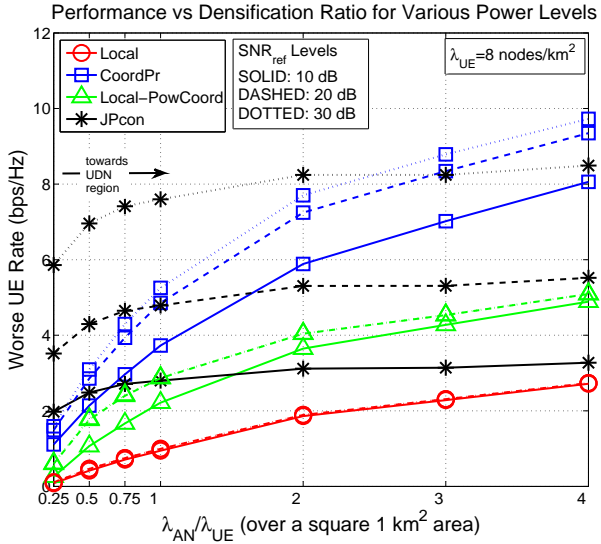}
\label{fig:RateVsDens_8UEs}}
\hfill
\subfloat[{Impact of UE density for various Densification Ratios}]
{\includegraphics[width=0.50\textwidth]{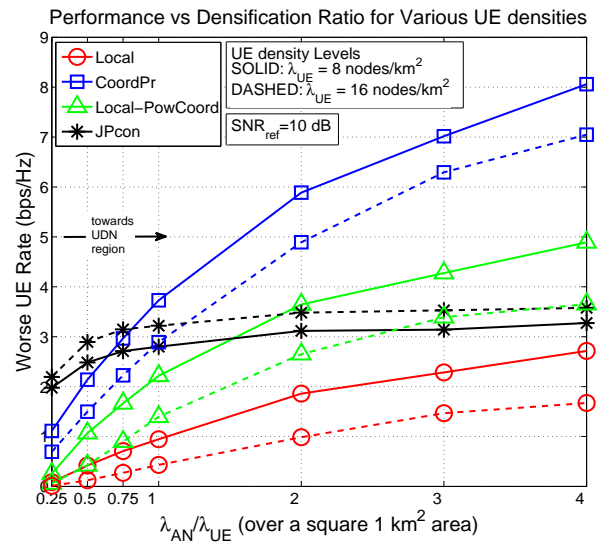}
\label{fig:RateVsDens_8and16UEs_pref_10dB}}
}
\caption{{Worse-UE Rate Performance various scenarios, system settings, and operation regions determined by $\lambda_{\text{AN}}/\lambda_{\text{UE}}, \text{ } p_{\text{ref}}, \text{ and } \lambda_{\text{UE}}$}.}
\label{fig:RateVsDens}
\end{figure*}

\textbf{2) Impact of Densification Ratio and Power Budget for constant UE density}: We now fix the UE density (to $\lambda_{UE}=8$ nodes/$\text{km}^2$) and examine the worse-UE rate scaling with an increasing AN density. As shown in Figure~\ref{fig:RateVsDens_8UEs} we start with regular non-dense setups (${{{\lambda _{{\text{AN}}}}} \mathord{\left/
 {\vphantom {{{\lambda _{{\text{AN}}}}} {{\lambda _{{\text{UE}}}}}}} \right.
 \kern-\nulldelimiterspace} {{\lambda _{{\text{UE}}}}}} < 1$) and move towards the ultra-dense network region for which the AN density even surpasses the UE density. Starting from the baseline uncoordinated ``Local" strategy, we observe a non-negligible worse-UE rate performance enhancement in the UDN region. Indeed, the strategy harnesses the densification gains in a two-fold way. First, each UE may be served by a closer AN, hence with an enhanced received signal power, while interference power from other active ANs remains more or less the same. Secondly, by increasing AN density, the number of co-AN UEs are reduced, hence more spatial dof can be utilized to increase per-UE performance (through maximum-ratio-transmission beamforming for example~\cite{BjJo13}). As far as the impact of power increase, we notice that it does not affect the baseline strategy (all red curves with circle markers coincide), since its uncoordinated nature always leads to an operation point lying in the interference-limited region.

The above densification-related gains are also harnessed by the two coordination strategies (2) and (3), as well. On top of them, ``Local−PowCoord" and ``CoordPr" achieve even higher rate performance due to their ability to control inter-UE interference not only locally (per-AN basis) but on a network-wise way. Moreover, an increase in available power seems to initially enhance performance, but from a point on its effect diminishes, since the operation point enters the interference-limited region. Note that this occurs later for ``CoordPr" due to its ability to jointly and optimally tune beamforming weights and the allocated power per UE, contrary to the decoupled and suboptimal approach followed by ``Local−PowCoord".

Interestingly, the ``JPcon" strategy exhibits a different behavior for varying densification ratio and reference SNR levels (power budgets). For heavier infrastructure density, negligible performance improvement levels are provided, due to the hard constraint on the cardinality of active ANs subset. We remind that for the current scenario of $\lambda_{UE}=8$ nodes/$\text{km}^2$ always the best size-2 ANs subset is selected for jointly serving all UEs, hence densification impact is rather limited. On the other hand, the diminishing effect of ever-increasing network power is not observed in this case. This is reasonable, since the employment of ``JPcon" resorts to a noise-limited network operation point. We argue that due to energy consumption considerations, future UDNs are expected to operate with limited power budgets, hence the coordination approaches may be preferable to the particular constrained cooperative processing strategy. Note also that for extreme densification ratios given by ${{{\lambda _{{\text{AN}}}}} \mathord{\left/
 {\vphantom {{{\lambda _{{\text{AN}}}}} {{\lambda _{{\text{UE}}}}}}} \right.
 \kern-\nulldelimiterspace} {{\lambda _{{\text{UE}}}}}} \gg 1$ and low power budgets (${\text{SNR}_{{\text{ref}}}}\sim10{\text{dB}}$), ``JPcon" will be even outperformed by the local (uncoordinated) strategy. Figure~\ref{fig:RateVsDens_8UEs} allows us to select the best strategy based on the densification and power availability conditions.

\textbf{3) Impact of UE density for various Densification Ratios}: Building upon the previous discussion, we also vary the UE density and explore its impact on the dependence of rate performance versus the densification ratio. Results for $\lambda_{UE}=8$ and $16$  nodes/$\text{km}^2$ (and fixed $\text{SNR}_\text{ref}=10$ dB) are presented in Figure~\ref{fig:RateVsDens_8and16UEs_pref_10dB}. As in the analysis for the proportionate AN and UE density increase in Figure~\ref{fig:SquareDens}, we again observe that for increasing UE population, the performance of the local and two coordination approaches deteriorate, contrary to the improvement of the cooperative strategy (JPcon). In addition, the crossing point between ``CoordPr" and ``JPcon" moves to the right towards the heavier UDN region (larger ${{{\lambda _{{\text{AN}}}}} \mathord{\left/
 {\vphantom {{{\lambda _{{\text{AN}}}}} {{\lambda _{{\text{UE}}}}}}} \right.
 \kern-\nulldelimiterspace} {{\lambda _{{\text{UE}}}}}}$ settings). In particular, for the 8 UE node density case, this lies below 0.75, whereas for the 16 UE node density case, surpasses 1. The dependence of rate performance on the UE population suggests that in order to support a set of rate guarantees on a per-UE basis, heavier network densification are required for larger UE node densities.

\section{Conclusion \& Future Work}\label{sec:Conclusion}
Extreme network densification empowered by efficient coordination capabilities is considered a strong candidate for fulfilling the highly challenging load requirements anticipated in future wireless networks. By intelligently exploiting the available spatial degrees of freedom in UDNs, significant performance enhancements can be achieved compared to uncoordinated network operation. In this paper we proposed a generic pairing optimization framework, which handles the association of users with access nodes and allows for great flexibility in harnessing the densification gains and limiting the induced overhead signaling. Leveraging this framework, we provided a detailed study on alternative solutions for operating such a UDN, and exposed various local, centralized or distributed coordinated, as well as cooperative spatial resources management strategies, exhibiting interesting trade-offs. We thoroughly explored the impact of critical operation parameters, namely density levels of AN and UE nodes as well as power budget constraints, revealed rate-scaling trends, and discussed the pros and cons of the various strategies. The current work may aid on the design and parametrization of efficient network coordination strategies applied to future 5G-UDNs.

For future work, we consider the study of involved overhead models that capture more accurately the requirements of each strategy in terms of exchanged signaling and data. Going beyond the cardinality-based approach followed in this paper, will allow us to better compare the various uncoordinated, coordinated and cooperative approaches. Other efficient spatial resources management strategies could be also explored, such as network clustering and partitioning. The impact of increasing the number of antennas per access node, following the massive-MIMO trend, worth also attendance, since it may lead to a drastic change in the interference environment, by allowing for creating highly directive spatial beams per user.

\section*{Acknowledgment}
This work has been performed in the context of the ART-COMP PE7(396)\textit{ ``Advanced Radio Access Techniques for Next Generation Cellular NetwOrks: The Multi-Site Coordination Paradigm"}, THALES-INTENTION and THALES-ENDECON research projects, within the framework of Operational Program "Education and Lifelong earning", co-financed by the European Social Fund (ESF) and the Greek State.

\appendix[Spatial Precoding Optimization Review]\label{sec:Appendix}
For the sake of completeness, we provide a brief review of the spatial precoding operation, based on \cite{ToPe09},\cite{BjJo13}, which follows the determination of AN-UE pairs described in Section~\ref{sec:CoordStrategies_Pairing}. Spatial precoding determines the beamforming weight vectors as well as the power-scaling, included in precoding vectors ${\left\{ {{{\mathbf{w}}_k}} \right\}}$, for every UE $k$. Regarding the problem of maximizing the minimum (worse) UE performance, the authors in \cite{ToPe09} have proposed a generic algorithm, leveraging (convex) second-order cone programming for detecting if a SINR level can be achieved or not by all UEs, and bisection search for locating the maximum achieved SINR (and equivalently rate) level. Given that the binary pairing variables $\left\{ {{\rho _{km}}} \right\}$ have been determined, we denote the single serving AN per UE $k$ as ${\mathcal{S}_k} = \left\{ {m \in \mathcal{M}:{\rho _{km}} = 1} \right\}$ (this holds true for a unique $m$ per user $k$), and the set of UEs sharing the AN $m$ as ${\mathcal{U}_m} = \left\{ {k \in \mathcal{K}:{\rho _{km}} = 1} \right\}$. We remind that $\mathcal{A}$ is the set of ANs serving at least one UE and that total power is evenly split among active ANs, thus ${{p_m \leq {p_{tot}}} \mathord{\left/
 {\vphantom {{p_m^{MAX} = {p_{tot}}} {\left| \mathcal{A} \right|}}} \right.
 \kern-\nulldelimiterspace} {\left| \mathcal{A} \right|}},\forall m \in \mathcal{A}$.
\paragraph{Feasibility check for a given target SINR level} Assuming an arbitrary SINR level $\theta _0^{\left( i \right)}$, which should be guaranteed for all UEs, feasibility could be checked by solving the SOCP problem formulated in \eqref{eq:PrecodingOptForm}. Note that the first two constraints sets are extracted by reordering the terms of the basic SINR per-UE expressions and exploiting the fact that phase-shifts do not affect optimal beamforming weights. The third constraints set guarantees that the maximum allocated power per active AN is not violated. At last, the fourth constraints set determines to which AN the data of each UE could not be available, by zeroing the corresponding precoding weight elements, depending on the followed strategy.
\begin{equation}\label{eq:PrecodingOptForm}
\small
\begin{gathered}
  {\text{find }}{\left\{ {{{\mathbf{w}}_k}} \right\}_{k \in K}}{\text{ }} \hfill \\
  {\text{subject to}} \hfill \\
  {\left\| {\begin{array}{*{20}{c}}
  1 \\
  {{{\left[ {\left( {{\mathbf{h}}_k^H{{\mathbf{w}}_1}} \right) \ldots \left( {{\mathbf{h}}_k^H{{\mathbf{w}}_K}} \right)} \right]}^H}}
\end{array}} \right\|_2} \leqslant \sqrt {1 + \frac{1}{{\theta _0^{\left( i \right)}}}} \Re \left\{ {{\mathbf{h}}_k^H{{\mathbf{w}}_k}} \right\}, \hfill \\
  \Im \left\{ {{\mathbf{h}}_k^H{{\mathbf{w}}_k}} \right\} = 0,\forall k, \hfill \\
  \sum\limits_{k \in {\mathcal{U}_m}} {{{\left\| {{{\mathbf{w}}_{km}}} \right\|}_2}}  \leqslant \sqrt {{{p_{tot}}} \mathord{\left/
 {\vphantom {{{p_{tot}}} {\left| \mathcal{A} \right|}}} \right.
 \kern-\nulldelimiterspace} {\left| \mathcal{A} \right|}} ,\forall m \in \mathcal{A} \hfill \\
  \forall k:\left\{ \begin{gathered}
  {\text{(Case I: Local and Coord) }}\forall m \notin {\mathcal{S}_k}:{{\mathbf{w}}_{mk}} = {{\mathbf{0}}_{L \times 1}} \hfill \\
  {\text{(Case II: JPcon) }}\forall m \notin \mathcal{A}:{{\mathbf{w}}_{mk}} = {{\mathbf{0}}_{L \times 1}} \hfill \\
\end{gathered}  \right. \hfill \\
\end{gathered}
\end{equation}
\paragraph{Search for the maximum feasible SINR level} Since the maximum achievable SINR level by all UEs is not a-priory known, a bisection search procedure could be applied for locating it. In particular, given an initial range for the SINR level $\left[ {\theta _0^{lb},\theta _0^{ub}} \right]$ and a convergence tolerance $\varepsilon$, the following routine returns the optimal SINR and the precoding weights per UE that can realize it.
\begin{algorithm}
\small
\caption {Bisection search for locating maximum common SINR}
\begin{algorithmic}[1]
\State Given $\left[ {\theta _0^{lb},\theta _0^{ub}} \right]$, convergence tolerance $\varepsilon$
\Repeat
    \State $\theta _0^{\left( i \right)} \leftarrow {{\left( {\theta _0^{lb} + \theta _0^{ub}} \right)} \mathord{\left/
 {\vphantom {{\left( {\theta _0^{lb} + \theta _0^{ub}} \right)} 2}} \right.
 \kern-\nulldelimiterspace} 2};$
    \State Solve SOCP feasibility problem in~\eqref{eq:PrecodingOptForm};
    \State If problem is feasible $\theta _0^{ub} \leftarrow \theta _0^{\left( i \right)}$ else $\theta _0^{lb} \leftarrow \theta _0^{\left( i \right)}$;
\Until $\left| {\theta _0^{ub} - \theta _0^{lb}} \right| \leqslant \varepsilon$.
\end{algorithmic}
\label{alg:BisectionSearch}
\end{algorithm}
\normalsize




\bibliographystyle{IEEEtran}
\bibliography{IEEEabrv,ARTCOMP-bib}

%\begin{thebibliography}{1}
%\end{thebibliography}

\end{document}